\documentclass[a4paper]{spie}  

\usepackage{amsmath,amsfonts,amssymb}
\usepackage{graphicx}
\usepackage[colorlinks=true, allcolors=blue]{hyperref}

\title{GIARPS: commissioning and first scientific results}

\author[a]{R. Claudi}
\author[a]{S. Benatti}
\author[a]{I. Carleo}
\author[b]{A. Ghedina}
\author[b]{J. Guerra}
\author[b]{F. Ghinassi}
\author[b]{A. Harutyunyan}
\author[c]{G. Micela}
\author[d]{E. Molinari}  
\author[e]{E. Oliva}
\author[e]{M. Rainer}
\author[e]{A. Tozzi}
\author[e]{C. Baffa} 
\author[a]{A. Baruffolo} 
\author[e]{V. Biliotti}
\author[f]{N. Buchschacher} 
\author[b]{M. Cecconi}
\author[b,i]{R. Cosentino}
\author[e]{G. Falcini} 
\author[a]{D. Fantinel} 
\author[e]{L. Fini}
\author[e]{E. Giani} 
\author[c]{E. Gonzalez--Alvarez} 
\author[b]{M. Gonzalez}
\author[b]{C. Gonzalez}
\author[a]{R. Gratton}
\author[b]{N. Hernandez}
\author[e]{M. Iuzzolino} 
\author[b]{M. Lodi}
\author[g]{L. Malavolta} 
\author[c]{J. Maldonado}
\author[h]{L. Origlia}
\author[e]{A. Puglisi}
\author[e]{N. Sanna}
\author[b]{J. San Juan} 
\author[i]{S. Scuderi} 
\author[j]{U. Seemann} 
\author[k]{A. Sozzetti}
\author[e]{M. Sozzi}
\author[b]{H. Perez Ventura} 
\author[b]{M. Hernandez Diaz} 
\author[b]{A. Galli}
\author[b]{C. Gonzalez} 
\author[b]{L. Riverol}
\author[b]{C. Riverol}
\affil[a]{INAF-- Astronomical Observatory of Padova, vicolo Osservatorio, 5, Padova, Italy}
\affil[b]{INAF -- Fundaci\'on Galileo Galilei, Rambla Jos\'e Ana Fern\'andez P\'erez, 7,  Bre$\tilde{n}$a Baja--TF, Spain}
\affil[c]{INAF-- Astronomical Observatory of Palermo, Piazza del Parlamento, 1, Palermo, Italy}
\affil[d]{INAF--Astronomical Observatory of Cagliari, Via della Scienza, 5, Cuccuru Angius, Selargius--Cagliari, Italy}
\affil[e]{INAF--Astrophysical Observatory of Arcetri, Largo Enrico Fermi,5, Firenze, Italy}
\affil[f]{Astronomy Department of Gen\'eve University, Chemin des Maillettes, 51, Versoix--Gen\'eve, Switzerland}
\affil[g]{Physics and Astronomy Depertment of Padova University, Via Francesco Marzolo, 8, Padova, Italy}
\affil[h]{INAF-- Astronomical Observatory of Bologna, Via Piero Gobetti, 93/3, Bologna, Italy}
\affil[i]{INAF--Astrophysical Observatory of Catania, Via S.Sofia 78, Catania, Italy}
\affil[j]{Institut für Astrophysik G$\ddot{\rm o}$ttingen, Georg-August Universit$\ddot{\rm a}$t, G$\ddot{\rm o}$ttingen, Germany}
\affil[k]{INAF--Astrophysical Observatory of Torino, Strada Osservatorio 20, Pino Torinese (TO), Italy}

\authorinfo{Send correspondence to R. Claudi\\R. Claudi: E-mail: riccardo.claudi@inaf.it}

\begin{document} 
\maketitle

\begin{abstract}
GIARPS (GIAno \& haRPS) is a project devoted to have on the same focal station of the Telescopio Nazionale Galileo (TNG) both high resolution spectrographs, HARPS--N (VIS) and GIANO--B (NIR), working simultaneously. This could be considered the first and unique worldwide instrument providing cross-dispersed echelle spectroscopy at a resolution of 50,000 in the NIR range and 115,000 in the VIS and over in a wide spectral range ($0.383 - 2.45\ \mu$m) in a single exposure. The science case is very broad, given the versatility of such an instrument and its large wavelength range. A number of outstanding science cases encompassing mainly extra-solar planet science starting from rocky planets search and hot Jupiters to atmosphere characterization can be considered. Furthermore both instruments can measure high precision radial velocities by means the simultaneous thorium technique (HARPS--N) and absorbing cell technique (GIANO--B) in a single exposure. Other science cases are also possible.
GIARPS, as a brand new observing mode of the TNG started after the moving of GIANO--A (fiber fed spectrograph) from Nasmyth--A to Nasmyth--B where it was re--born as GIANO--B (no more fiber feed spectrograph). The official Commissioning finished on March 2017 and then it was offered to the community. Despite the work is not finished yet. In this paper we describe the preliminary scientific results obtained with GIANO--B and GIARPS observing mode with data taken during commissioning and first open time observations.
\end{abstract}


\section{Introduction} \label{sec:intro}
The discovery of a hot Jupiter orbiting 51Peg\cite{mayorandqueloz1995} triggered the quest for extrasolar planets using the radial velocity technique, by which the presence of a planetary companion is inferred by the wobble it induces on the parent star. In the present, the search for more exotic stellar hosts and the race for the lightest planets point towards M dwarfs. These stars, which are the most abundant in the Universe, are also the smallest ones. Since the RV variation induced by a planet on a star scales with M$_{\star}^{-2/3}$, the amplitude of the effect induced on a M star is significantly larger. As an example, a planet of an identical mass at the same distance from the stellar host produces a RV variation with an amplitude $\sim 3$ times larger on an M5 star than on a G2 star. The drawback is that since they are much colder, M dwarfs are much fainter in optical wavelengths. The RV surveys of low-mass stars points then towards the exploration of a new wavelength domain, the IR, where the luminosity of these objects peaks. This type of stars can show surface inhomogeneities like stellar spots, being young and active, which can mimic or hide Doppler signal due to a planet. Observing in the NIR, as opposed to VIS, the contrast between these surface inhomogeneities and the stellar disk is strongly reduced\cite{reinersetal2010},  helping to discriminate between colored signal (activities, pulsations etc.) and planetary signal. This highlights another advantage of measuring radial velocities in NIR.

In this framework GIARPS (GIAno \& haRPS--N) \cite{claudietal2016}, the new common feeding for both the high resolution spectrographs, HARPS-N\cite{cosentino2014}  in the visible and GIANO\cite{oliva2006} in the NIR, represents a good chance to investigate this class of objects in the next future. GIARPS allows to have the two instruments on the same focal station of the Telescopio Nazionale Galileo (TNG) working simultaneously.  

The science case is very broad, given the versatility of such an instrument and the large wavelength range, encompassing mainly extra-solar planet science starting from rocky planet search and hot Jupiters atmosphere characterization can be considered. But not only, also young stars and proto--planetary disks, cool stars and stellar populations, moving minor bodies in the solar system, bursting young stellar objects, cataclysmic variables and X-ray binary transients in our Galaxy, supernovae up to gamma--ray bursts in the very distant and young Universe, can take advantage of the unicity of this facility both in terms of contemporaneous wide wavelength range and high resolution spectroscopy.

\section{WOW and GAPS\ 2.0} \label{sec:wow}
WOW (A Way to Other Worlds) is a program, funded in the framework of the funding scheme {\it Progetti Premiali} of the Italian Ministry of Education, University, and Research, that aims to take in cooperation and connect the Italian extrasolar planets community through the realization of several projects, all aimed towards the following main points:
\begin{enumerate}
\item The full exploitation of the planetary data obtained by the Italian community on national and international facilities.
\item The discovery of a few of rocky planets around low mass stars, in the habitable zone, with the Italian Telescope TNG.
\item An initial statistics of exoplanets (of various mass and temperature) in the solar neighbourhood and the estimate of their capability to host life.
\item A determination of the expected physical properties of exoplanetary atmospheres that will be observed with future planned instruments.
\item A {\it complete} database of the conditions that can be found in planetary atmospheres of our Solar System.
\item The definition of what is a biomarker.
\item Feasibility studies to improve the performance of already working facilities.
\item The definition of scientific requirements for the future instrumentations to be transformed in engineering requirements.
\item Feasibility analysis of new concepts for innovative instrumentations.
\end{enumerate}
Last three points allowed to fund the refurbishment of GIANO based on the scientific requirements defined by a growing community that is gathered in the GAPS (Global Architecture of Planetary Systems) project. At the beginning (2012) GAPS Project, the collaboration among part of the Italian exoplanetary community, planned to exploit the great capabilities of the HARPS--N instrument. It gathers more than 60 astronomers from several institutes of the Italian National Institute for Astrophysics (INAF) and Italian Universities (Padova, Torino and Milano). A technical and scientific support is also provided by a few collaborators from European and American Institutions. The main purpose of GAPS is the study and the characterization of the architectural properties of planetary systems through the radial velocity technique, by analyzing the distributions of planetary parameters and their correlations with those of the host star. After five years of HARPS--N observations and analysis, GAPS have developed an optimized observing strategy and new analysis tools, in particular for those objects which require many data and specific treatment of the RV time series. Anyway, new perspectives are foreseen for GAPS using GIARPS that is now available at TNG with high impact on the exoplanetary research. The extension of the wavelength range will open to the GAPS community new scenarios and objectives in the study of the extra- solar planets. The now called GAPS 2.0 project is now focused on the study of a number of hot Jupiter atmospheres and the detection/confirmation of planets around a sample of young stars. GAPS has obtained the status of long program with GIARPS to start its new scientific programs.

\section{GIARPS}
\begin{figure} [ht]
   \begin{center}
   \begin{tabular}{c} 
   \includegraphics[height=7cm]{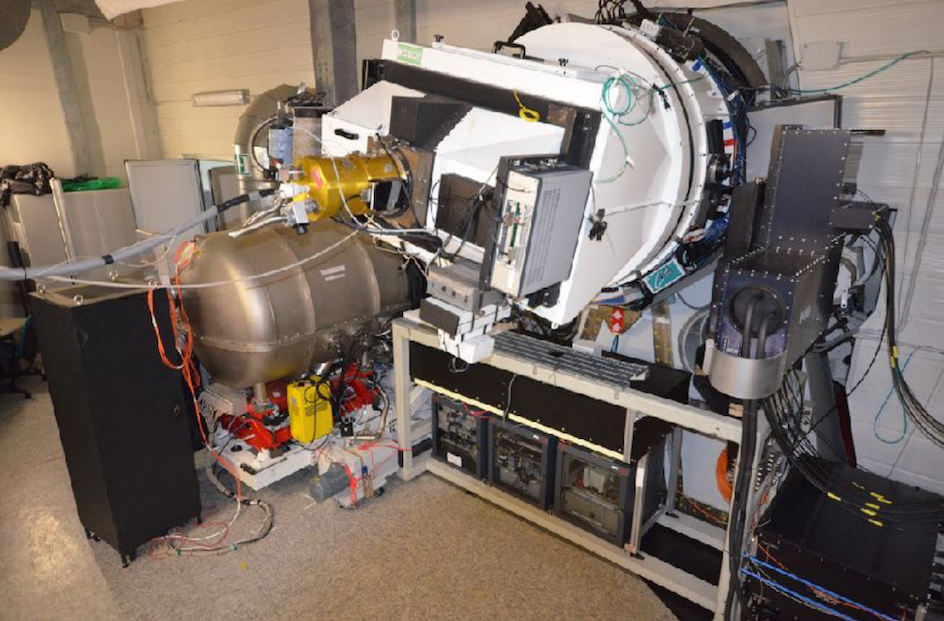}
	\end{tabular}
	\end{center}
   \caption[example] 
   { \label{fig:giarps} Picture of the current configuration of the Nasmyth-B focus of TNG. \textit{Left}: the electronics and the dewar containing GIANO-B, mounted on a dedicated platform. \textit{Top center}: the Low Resolution Spectrograph (LRS) that hosts the entrance slider where is located the dichroic. \textit{Bottom centre}: the new GIANO-B preslit that receives the light through LRS. It is mounted inside the lodgment of SARG (the previous high-resolution spectrograph of TNG) together with the corresponding electronics. \textit{Right}: The front-end and calibration units of HARPS-N. }
\end{figure}

As mentioned in Sect. \ref{sec:wow}, in 2014 the coordinators of the \textit{WOW} Progetto Premiale proposed a full exploitation of the two high resolution spectrographs of TNG, HARPS-N and GIANO, aiming to obtain high resolution spectroscopy and high precision radial velocities in both visible and near-infrared ranges. HARPS-N is a cross-dispersed echelle spectrograph, mounted at TNG in 2012 at the Nasmyth-B focus. It works in the visible (VIS) range between $0.39$ and $0.68$ $\mu$m with a spectral resolution of R $=115,000$. HARPS-N operates in an extremely stabilized and monitored environment which allows to reach the extraordinary radial velocity precision of 1 m\,s$^{-1}$ or less. 
GIANO  is a high-resolution (R $=50,000$) near-infrared (NIR) echelle spectrograph working from 0.95 to 2.45 $\mu$m. Despite it was designed to be mounted at the Nasmyth-B focus of the TNG and directly fed by the telescope, scheduling issues in 2012 compelled the builders to mount it in the other focus chamber (Nasmyth-A). This implied the use of  ZBLAN fibers to feed the spectrograph, which introduced a modal noise  in the spectra that degraded significantly the instrument efficiency\cite{origlia2014}. With the proposal to use the two instruments simultaneously, after a feasibility study and the definition of a new pre-slit system\cite{tozzi2016}, GIANO was finally moved at the end of 2016 to the Nasmyth-B chamber, fixed on a dedicated platform attached to the telescope fork and aligned with the new pre-slit (Figure \ref{fig:giarps}).
After this operation, the instrument, now called ``GIANO-B", has been able to work with HARPS-N thanks to the insertion of a dichroic in the light path that splits the VIS and NIR beams to the corresponding spectrograph. The dichroic is located on an entrance slider that allows to choose the preferred observing mode: HARPS-N only, GIANO-B only and GIARPS. More details on the instruments are provided in Ref.\ \citenum{claudietal2016}.

\section{Commissioning Results} \label{sec:commissioning}
The upgrade of GIANO started with the design of a preslit optics\cite{tozzi2016} able to feed GIANO without the use of optical fibers. The validation of the GIARPS preslit and of the new configuration at the Nasmyth--B focal station of the TNG was conducted in three different commissioing runs. 

In the first commissioning run (see Sec. \ref{sec:preslitcomm}), the preslit was integrated and aligned and the GAG (GIANO--B auto guide) tested and validated. The second commissioning aimed to test the new GIANO--B configuration, the calibration units and the scheduler, sequencer, preslit, autoguide softwares (Sec. \ref{sec:gianobcomm}). Finally, the third and last commissioning run aimed to optimize the GIANO--B performances and test the GIARPS mode, with the insertion of a dichroic (Sec. \ref{sec:giarpscomm}).

\subsection{Preslit and Auto Guide commissioning}
\label{sec:preslitcomm}
The first commissioning run has been spent to align the preslit and to check the pupil position, while its dimensions were checked for different elevation values of the telescope in daytime with artificial lights and different star targets in night-time. 
The GAG field of view (FoV) was defined observing a bright star, like Vega, and moving it in the FoV, while the GAG pixel scale has been determined with two different methods: through the telescope off-sets and observing binary systems with known distance. The two measurements are consistent, the nominal value of 0.028 arcsec/pixel while the measured FoV is $20 \times 20$ arcsec$^2$.

\begin{figure} [ht]
   \begin{center}
   \begin{tabular}{c} 
   \includegraphics[height=7cm]{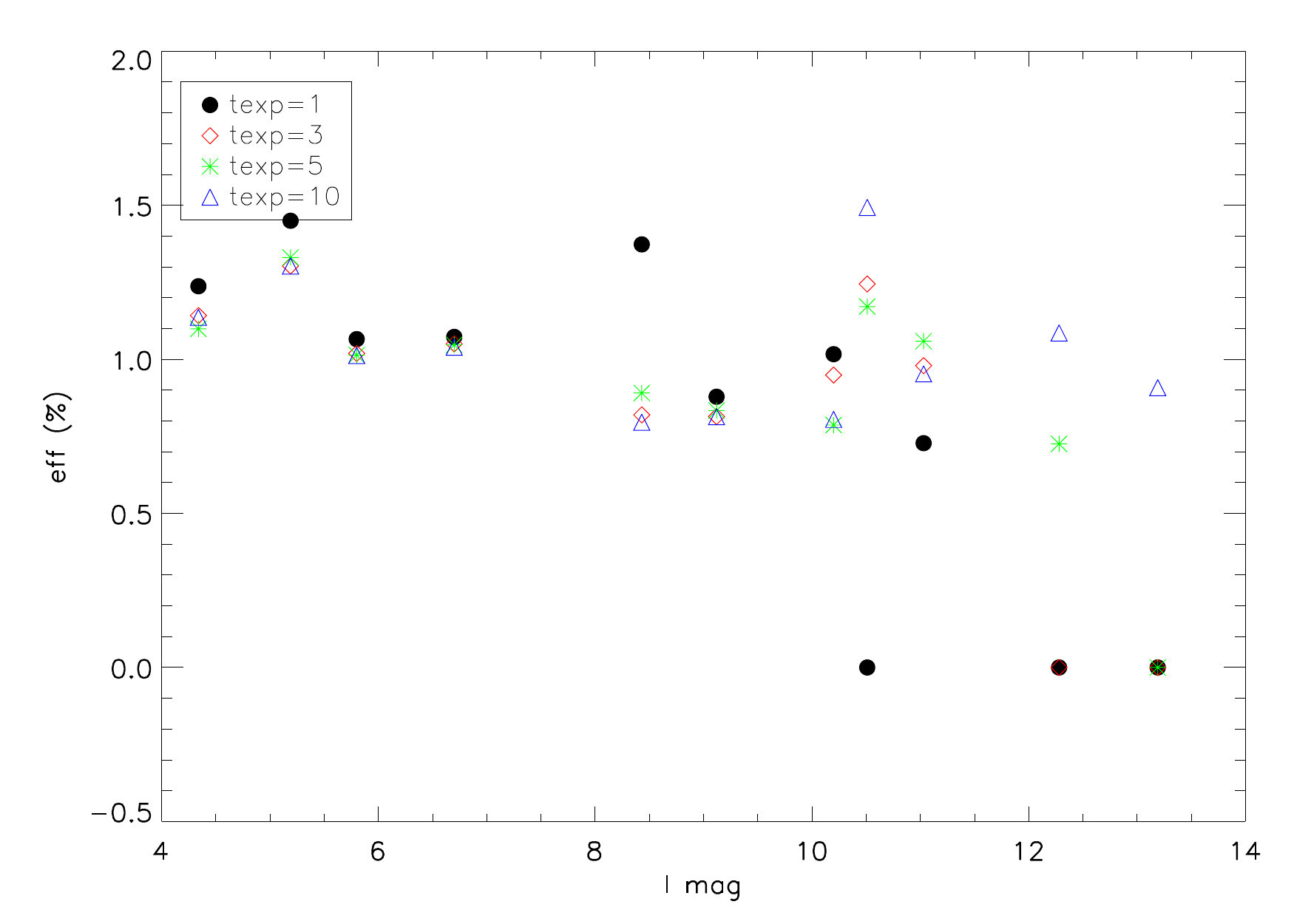}
	\end{tabular}
	\end{center}
   \caption[example] 
   { \label{fig:preslit_eff} 
Efficiency (in percentage) of the preslit as function of I magnitude at different exposure times, without filters.}
   \end{figure}

In order to define the correct exposure times for the AG as function of the stellar magnitude, we perform several observations with the GAG camera of photometric standard stars ranging between I= (4 - 13)\ mag with several exposure times.
These measurements were performed both with and without the insertion of neutral filters with different optical density, allowing to pass the 4.5\%, 0.33\%, and 0.02\% of the light, useful to avoid that the brighter objects saturate the detector of the Camera. A table of stellar magnitudes with the related exposure times has been constructed, observing photometric standard stars with different magnitudes. 
At this point the photometric efficiency (including atmosphere, telescope and Quantum Efficiency of the CCD) of the preslit has been calculated at different exposure times, resulting around 1\% as showed in the Fig. \ref{fig:preslit_eff}. 
Finally, we determined the limit I magnitude for the guiding camera, that results to be 12 (without filters). A new version of the GAG software able to binning the detector of the camera, has been released in May 2018. The photometric efficiency of GAG was checked and a new table of stellar magnitudes with the related exposure times was recompiled. In this new version, the camera gain about two magnitudes deeper in the useful magnitude for guiding or to be precise the limit magnitude (texp=10s, s/n@peak$\sim 14$, seeing$\sim 0.75$\ arcsec, at z=40deg) is z'$\sim 14$. 

\subsection{GIANO-B commissioning}
\label{sec:gianobcomm}

\begin{figure} [ht]
   \begin{center}
   \begin{tabular}{c} 
   \includegraphics[height=7cm]{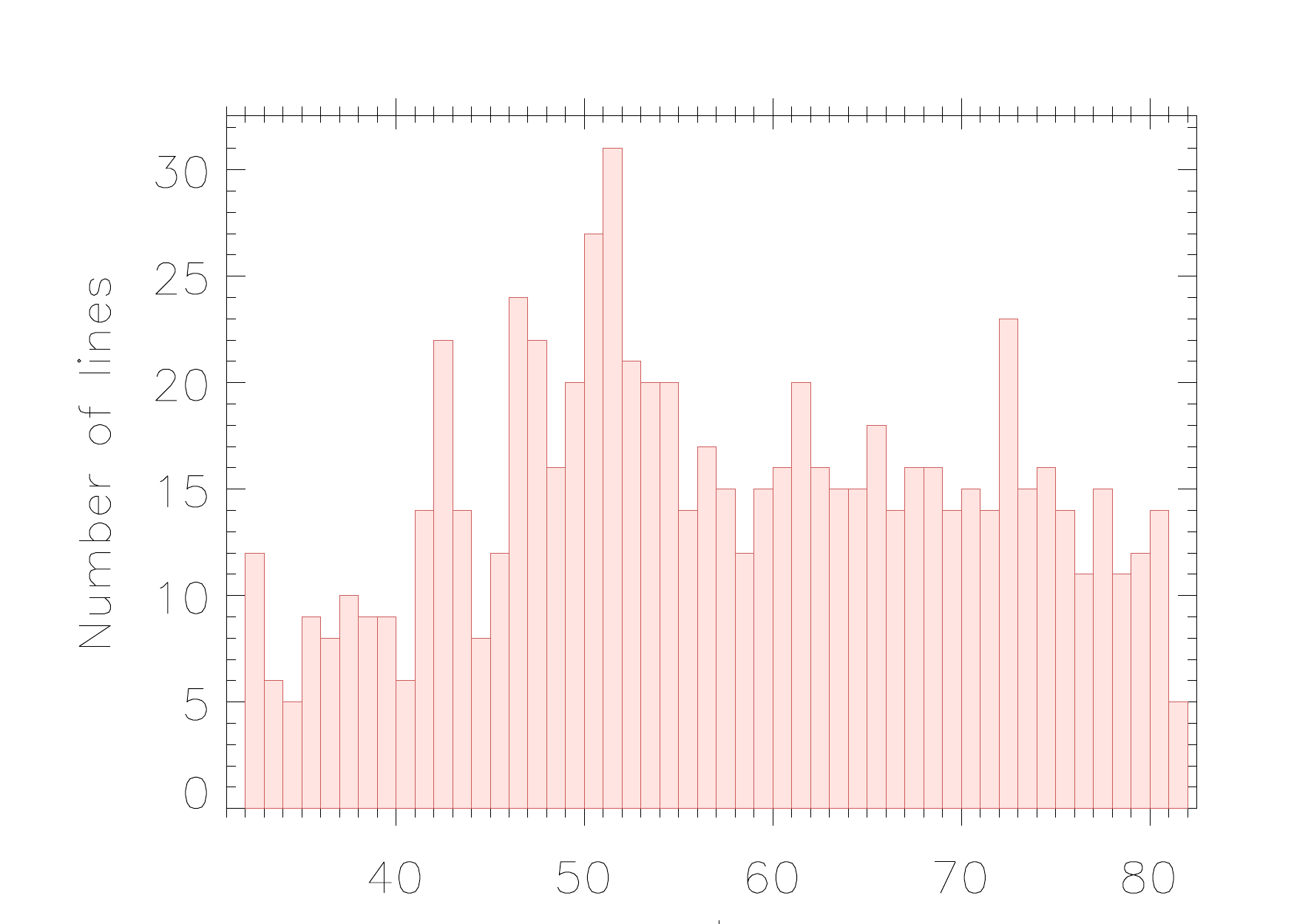}
	\end{tabular}
	\end{center}
   \caption[example] 
   { \label{fig:nlines} 
Number of U-Ne lines for each order.}
   \end{figure}

\begin{figure} [ht]
   \begin{center}
   \begin{tabular}{c} 
   \includegraphics[height=7cm]{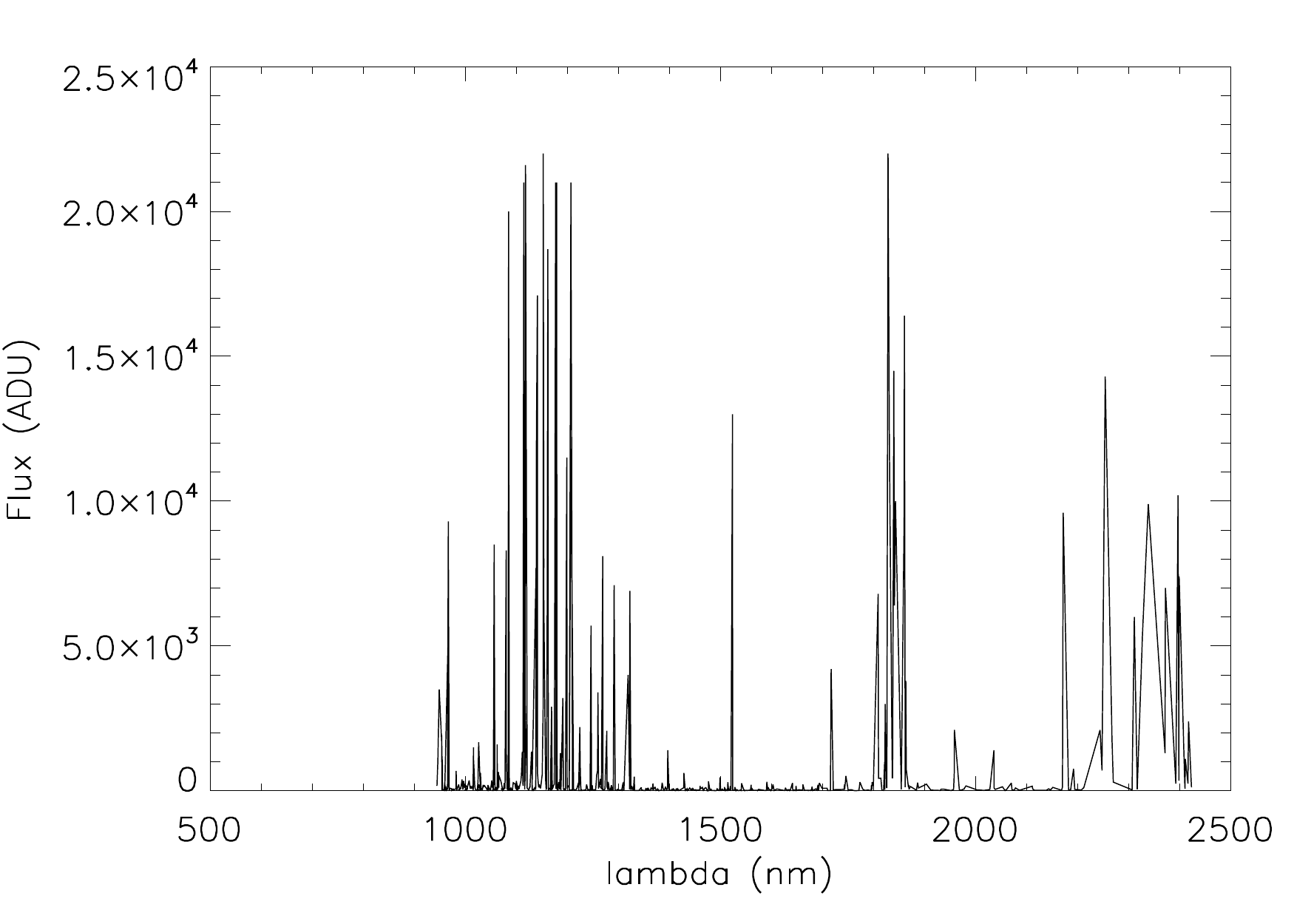}
	\end{tabular}
	\end{center}
   \caption[example] 
   { \label{fig:linesflux} 
Flux of the U-Ne lamp's lines (ADU in 30 seconds) as function of the wavelength.}
   \end{figure}

This phase consisted first in the alignment of the instrument with the rest of the system, including the slit. Calibration frames has been acquired in order to estimate the height of the slit of GIANO-B through the flat-field lamp and to check the stability of the U-Ne lamp.
Fig. \ref{fig:nlines} shows the number of lines for each order of the U-Ne spectrum, while Fig. \ref{fig:linesflux} shows the flux of the U-Ne lines as function of wavelength.

First observations with GIANO-B were performed in order to clock the overheads  during an AB nodding exposure (see Section \ref{sec:gofio}), resulting to be about 3 minutes, and to test the sequencer software and observation templates\cite{harutyunyan2018}.  

\subsection{GIARPS commissioning}
\label{sec:giarpscomm}
\begin{figure} [ht]
   \begin{center}
   \begin{tabular}{c} 
   \includegraphics[height=7cm]{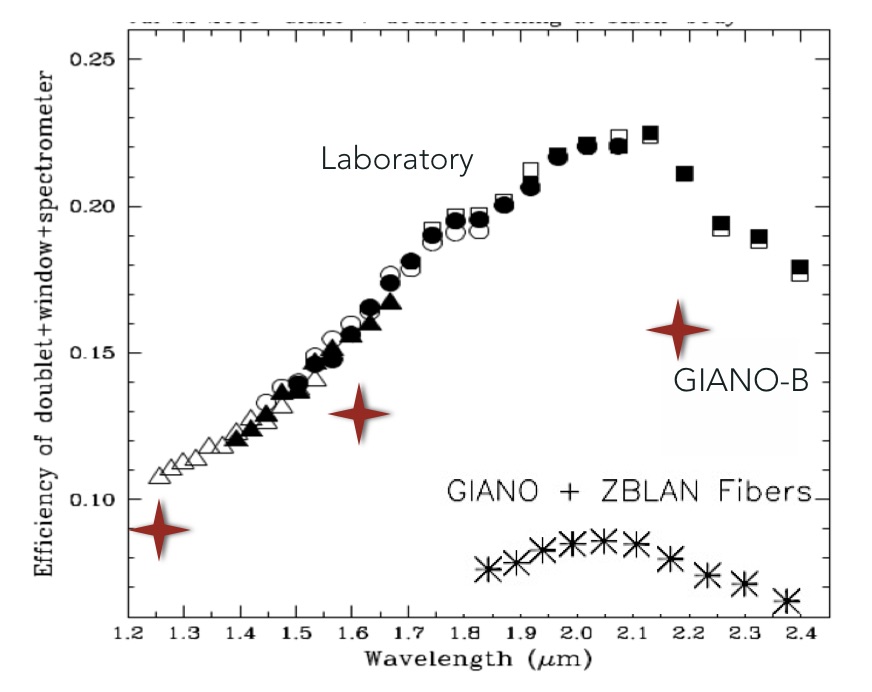}
	\end{tabular}
	\end{center}
   \caption[example] 
   { \label{fig:effi} 
The measured efficiency of GIANO--B (red stars) compared with the efficiency measured in the laboratory and the efficiency of GIANO--A feeded by the fibers.}
   \end{figure}

During this last commissioning the activities regarded the optimization of GIANO-B configuration, the sequencer and the autoguide final test, and the test of observations in GIARPS mode after mounting the dichroic. 
The focus and pointing model tests for GIANO-B have been performed. Furthermore,  in order to measure the actual efficiency of GIANO-B, without the loss of stellar flux due to the limited size of the slit and in an independent way by the seeing, we observe a standard star without the GIANO slit.  The calculated efficiency is 15\% in the K band (see Figure\ \ref{fig:effi}).

After the dichroic was mounted and aligned to the system, we performed the first GIARPS-mode observations, testing the focus of both the instruments with the dichroic. Further, we tested the new NSTS (New Short Time Scheduler) software adapted for GIANO--B and the GIANO-B sequencer (the commands dispatcher). We prepare and submit the observing templates necessary to perform both nodding and stare observing modes and test the performances of the system.  At the end of the exposure, when the spectrum is stored in the disk, the new data reduction pipeline, GOFIO\cite{rainer2018}(see also Sec. \ref{sec:gofio}) start to do its job. Details on the software that manages the creation of rough data files and coordinates the storing of data files into the TNG archive and the automatic start of GOFIO are in Ref.\ \citenum{harutyunyan2018}.

\subsection{The GOFIO Pipeline}\label{sec:gofio}
The different feeding of the GIANO-B spectrograph, from fibers to long slit, implied a different method to extract the spectra and provided the opportunity to create a dedicated data reduction pipeline. The GIANO-B reduction software, ``GOFIO" \cite{rainer2018}, processes the calibrations as well as all the spectra in real time during the observing night, obtained with the nodding or stare acquisition modes\footnote{According to the scientific target to observe it is possible to select a specific type of acquisition, aiming to properly remove the thermal background which is crucial for infrared instruments. In the nodding mode (mostly used for point sources) the target is alternatively observed in two well defined positions of the slit and subsequently subtracted one to each other. Similar method is used for extended objects, where the target is positioned in the center of the slit, which is entirely illuminated. In this case the background subtraction is performed by using sky spectra acquired close to the region of the target. GOFIO reduces both the acquisition modes.}.
GOFIO is part of the full online Data Reduction Software \cite{harutyunyan2018} that manages the data flow from the observation to the archive. It has been successfully tested during the second and third GIARPS Commissioning, providing the extracted spectra used for instrument characterization and scientific validation. 
An offline version of GOFIO will be available in the near future for GIANO-B/GIARPS users, allowing custom data reduction according to the scientific requirements of the observer and, if needed, to compute the radial velocity using telluric lines as a reference system (see Sect. \ref{sec:rv}).

\section{Radial Velocity Measurements} \label{sec:rv}
The method to extract radial velocities (RV) from GIANO and GIANO--B spectra follows the approach described in Carleo et al. 2016 \cite{carleo2016}: the telluric lines are used as wavelength reference and the Cross-Correlation function method is used to determine the stellar RV. For this purpose, we start with the preparation of files including evaluation of the correction to the barycenter of the solar system and a resampling of spectra in order to have a constant step of 200 $ms^{-1}$ in RV.  
 Then the spectra are normalized by dividing them to a fiducial continuum.
One of the most important steps for NIR RVs is the removal of the telluric lines from science spectra. For this purpose, we create the best telluric spectrum from a library including several telluric spectra acquired over time. This best telluric spectrum is shifted to take into account small shifts in the dispersion zeropints, and scaled in intensity to take into account the different airmass between telluric and science spectra. Finally, the obtained telluric spectrum is subtracted from the science spectra. This approach, that determines the best shift and scale parameters, allows to minimize the RMS of the subtracted spectra. \\
From the best telluric spectrum and the cleaned stellar spectra we construct the digital telluric and stellar masks respectively. Then, the procedure performs the cross correlation of individual orders of the normalized spectra with these masks (both stellar and telluric), with derivation of individual CCF and RVs. Through the weighted sum of the individual CCFs we obtain the stellar and the telluric RVs: the latter are finally subtracted from the former, providing the final relative stellar RVs. The uncertainties are then evaluated taking into account the photon statistics. \\
Anyway, a slightly different approach is considered. It takes into account the weight of the single orders, some of them being affected by telluric lines and thus contributing in different ways to the determination of the RVs. Starting from the RV of individual orders, we calculate the weighted average RV for each exposure and its corresponding error. As a final step, we derive the bisector velocity span (BIS \cite{carleo2016}) of the CCF and we calculate the uncertainties on this quantity by considering the fractions of the CCF used for the derivation of the BIS, resulting in $\sqrt{10}{\sigma_{RV}}$, where $\sigma_{RV}$ is the RV error.\\
This code will be implemented in GOFIO pipeline \cite{rainer2018} in order to have a completely automated computation procedure. 

\section{First Scientific Results}
\subsection{HD~3765}
\begin{figure} [ht]
   \begin{center}
   \begin{tabular}{c} 
   \includegraphics[height=7cm]{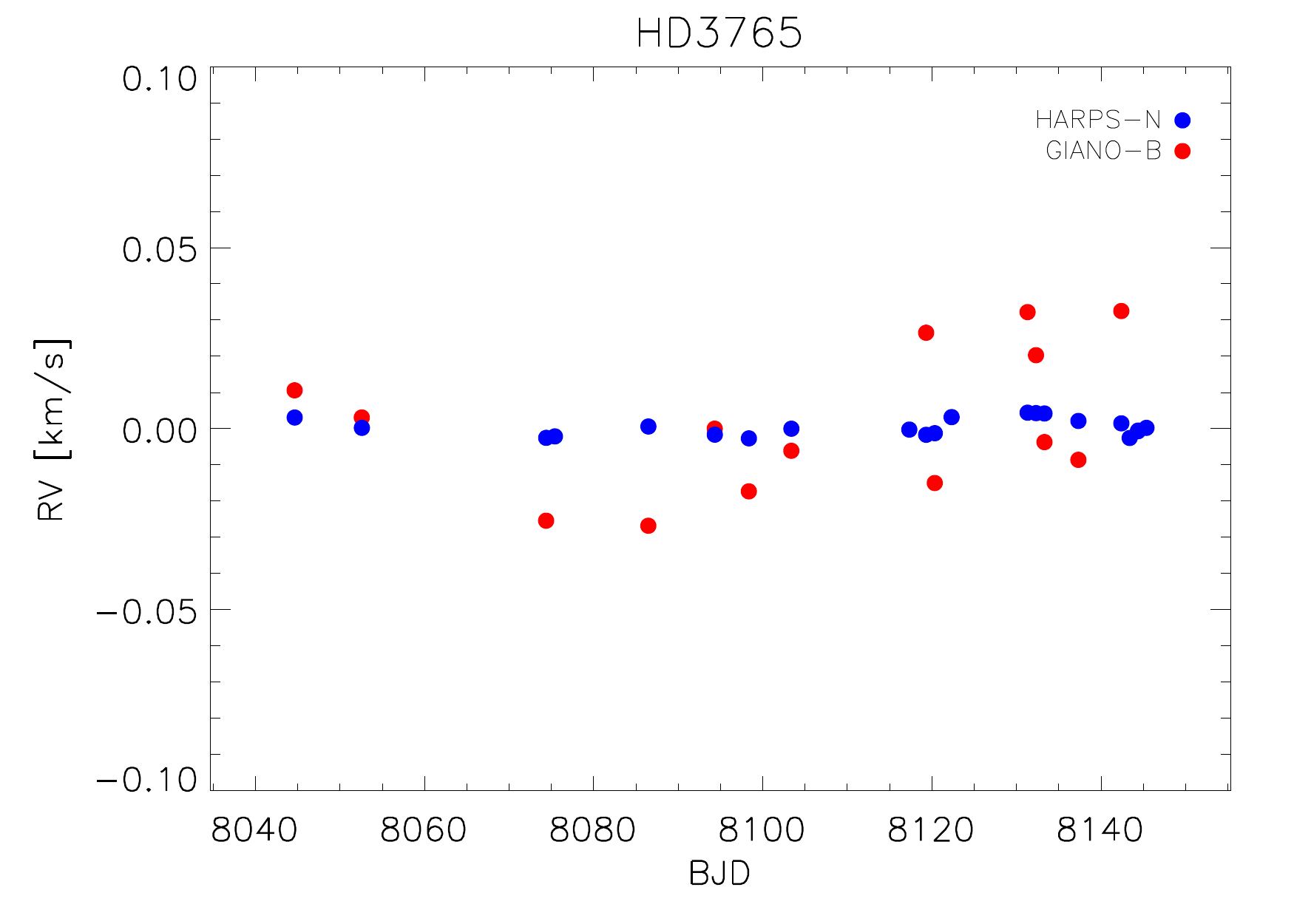}
	\end{tabular}
	\end{center}
   \caption[example] 
   { \label{fig:HD3765} 
Radial velocities of the standard star HD3765 from GIANO-B (red points) and HARPS-N (blue points) spectra.}
   \end{figure}

HD~3765 is a K2V star with magnitudes J=5.69, H=5.27, and K=5.16 \cite{cutri2003}. It has a low activity level \cite{Strassmeier2000,Martinez-Arnaiz2010} and a constant RV (RV jitter of 2.4 $ms^{-1}$) \cite{batten1983,isaacson2010} and was then used as a standard for testing instrument performances and to determine the GIANO-B stability.
We observed HD~3765 over one semester in GIARPS mode: the NIR spectra have a typical SNR values of 163 and internal errors in RVs of 25 $ms^{-1}$ (see Tab. \ref{tab:stars}), while the VIS counterpart a typical SNR of 200 and internal errors of 0.3 $ms^{-1}$ Fig. \ref{fig:HD3765} shows the resulting RVs both in visible (blue points) and near-infrared (red points) ranges. While the VIS RVs present a rms scatter of 2.3 $ms^{-1}$, the NIR RVs have an rms scatter of 8 $ms^{-1}$ over few days and 14 $ms^{-1}$ over the whole semester, which represent the short-term and long-term stability, respectively.

\subsection{Combined VIS-NIR observations to confirm or retreat planet candidates}
The very first result obtained by using data from the GIARPS Commissioning is the retreat of a controversial exoplanet around the K5 young star, BD+20\,1790\cite{carleo2018}, member of AB Dor moving group (age $\sim 150$ Myr). A significant RV modulation in the VIS range (having semi-amplitude R $\simeq$ 1 km s$^{-1}$) has been interpreted as a signature of a massive hot Jupiter with a period of 7.8 days \cite{HO2010,HO2015}. Anyway, doubts on the real presence of this companion have been raised\cite{Figueira2010}, attributing the RV variations to photospheric processes, since the RV modulation was incompatible with the proposed planet and because of the strong correlation between RV and the bisector span (a robust stellar activity indicator).  
Being BD+20\,1790 a very active star ($\log R'_{HK}=-3.7$), the combination of high-resolution spectroscopy in the VIS and NIR allows in principle to discriminate between Keplerian motion (due to a low-mass companion) and astrophysical noise, since the former is achromatic with respect to the latter. Actually, RV modulations in the NIR range can be significantly reduced\cite{Prato2008,Mahmud2011,Crockett2012}. For this reason we compare VIS archive data of BD+20\,1790 with high precision RVs obtained with GIANO/GIANO -- B and IGRINS\cite{Mace2016} spectrographs. 
   \begin{figure} [ht]
   \begin{center}
   \begin{tabular}{c} 
   \includegraphics[height=12cm,angle=270]{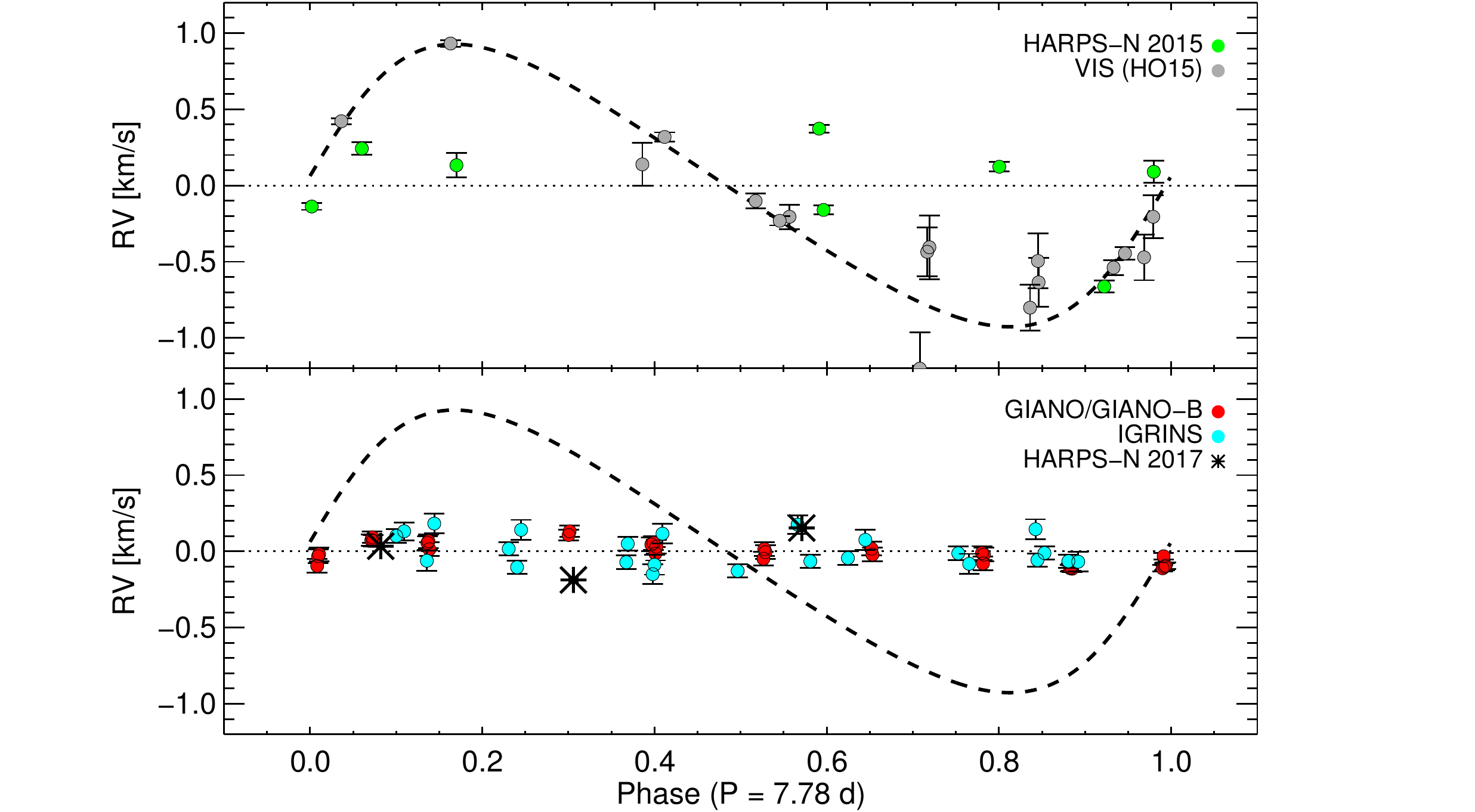}
	\end{tabular}
	\end{center}
   \caption[example] 
   { \label{fig:bd} 
\textit{Top panel}: Orbital fit (black dashed line) obtained with the visible data (FOCES, SARG, and HERMES RVs\cite{HO2010,HO2015}, grey dots) and HARPS-N 2015 RVs (green dots), overplotted. \textit{Bottom panel}: Orbital fit (black dashed line), GIANO/GIANO-B (red dots), IGRINS (light blue dots), and HARPS-N 2017 (black asterisks, two acquired in GIARPS mode) RVs\cite{carleo2018}.}
   \end{figure} 
As showed in Fig. \ref{fig:bd} the average RV amplitude in the NIR is significantly lower then VIS data, as expected when the RV jitter is due to stellar activity: the claimed massive planet around BD+20~1790 is therefore ruled out by our data. In this case it is showed the crucial role of multi-wavelength spectroscopy when observing young active stars and the opportunity to investigate key issues for the evolution of planetary systems, since we excluded the only previously known case of a hot Jupiter orbiting a star between 20 to 200 Myr old. Further investigations are needed to achieve a firm conclusion on this and many other astrophysical questions, but thanks to facilities like GIARPS,  providing simultaneous observations in VIS and NIR bands, this method can reach its maximum potential. \\

\begin{figure} [ht]
   \begin{center}
   \begin{tabular}{c} 
   \includegraphics[height=10cm, trim=0 26cm 0 0]{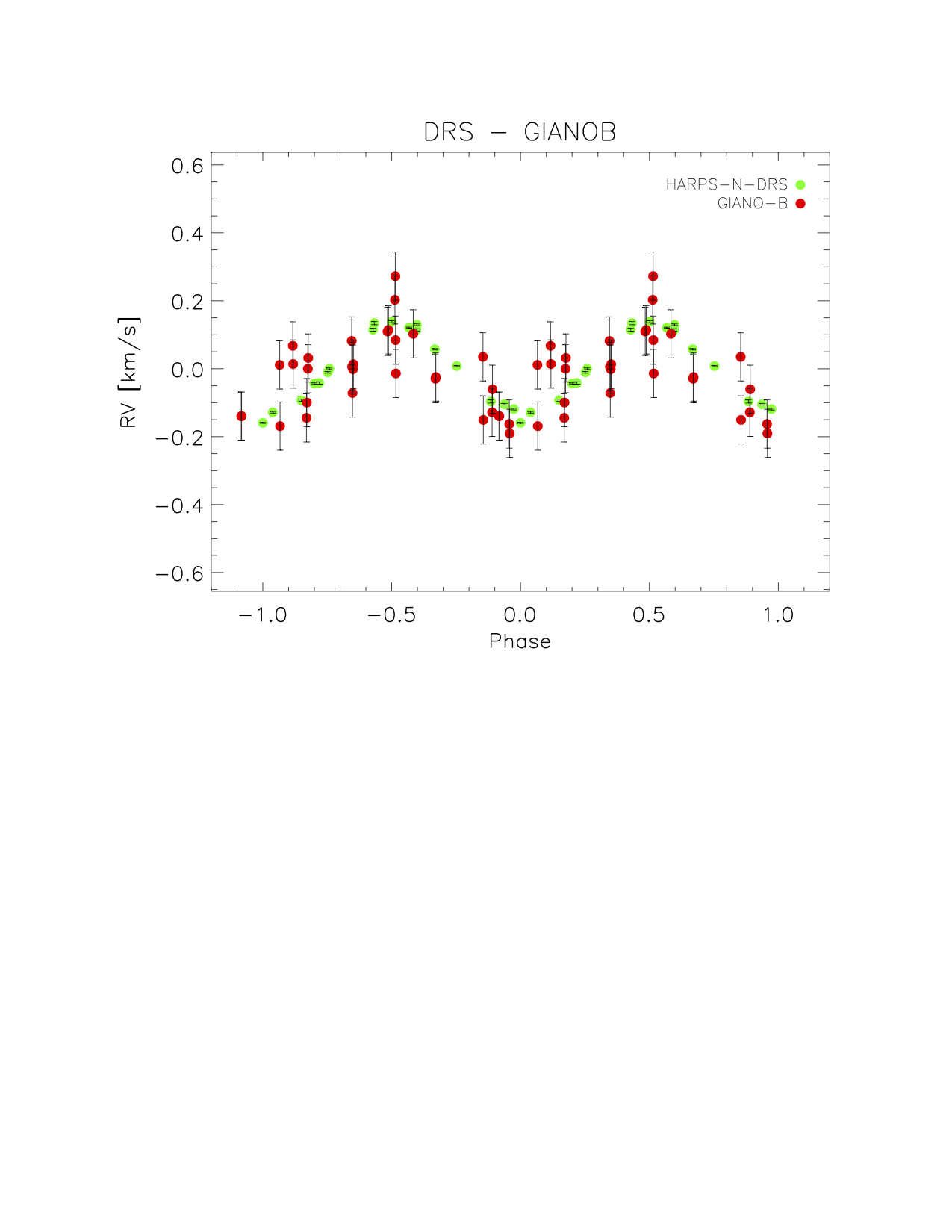}
	\end{tabular}
	\end{center}
   \caption[example] 
   { \label{fig:YO13} 
Radial velocities of YO13 from GIANO-B (red points) and HARPS-N (green points) spectra, phase-folded with the orbital period.}
   \end{figure}

As an opposite example to BD+20~1790, we present the case of a K4 star with H magnitude of 7.8, included in the GAPS\ 2.0 Young Object sample with code name ``YO13". It was selected after the claim of an hot Jupiter as part of the GAPS Pilot program during the TNG Observation period 36, when the limited allocated time prevented a blind planet search to the benefit of a monitoring of planet candidates. The radial velocities measured from GIANO-B (red points) and HARPS-N (green points) simultaneously (GIARPS mode) plotted in Fig. \ref{fig:YO13} are phase-folded to the orbital period. Both datasets show the same amplitude in the RV modulation and a similar rms scatter, leading to the confirmation of the claimed hot Jupiter. However, a thorough analysis of this case is ongoing.

\subsection{Internal errors and RV Precision}
In order to understand the GIANO-B performances and to assess the precisions of RV measurements, we analyzed different stars with different magnitudes (two RV standard stars and four targets of the GAPS 2.0 YO sample). The characteristics of the stars of our sample are reported in Tab. \ref{tab:stars}. 

\begin{table}[ht]
\caption{List of stars of our sample together with the H magnitude, the SNR value, $vsini$, internal error and r.m.s scatter of RVs obtained from our analysis.} 
\label{tab:stars}
\begin{center}       
\begin{tabular}{|l|c|c|c|c|c|} 
\hline
\rule[-1ex]{0pt}{3.5ex}  Star & Hmag & SNR & vsini & Internal error & RV r.m.s. scatter  \\
\rule[-1ex]{0pt}{3.5ex}       &     &      & $kms^{-1}$ & $ms^{-1}$ & $ms^{-1}$  \\
\hline
\rule[-1ex]{0pt}{3.5ex}  YO01 & 8.6 & 27   & 30.5 & 769        & 946   \\
\hline
\rule[-1ex]{0pt}{3.5ex}  YO13 & 7.8 & 45   & 3.0  & 71         & 114  \\
\hline
\rule[-1ex]{0pt}{3.5ex}  YO14 & 7.3 & 75   & 23.8 & 546 & $-$  \\
\hline
\rule[-1ex]{0pt}{3.5ex}  YO17 & 4.8 & 163  & 3.0 & 15 & 17  \\
\hline 
\rule[-1ex]{0pt}{3.5ex}  HD~3765 & 5.3 & 163  & 1.7 & 25 & 8 - 14  \\
\hline 
\rule[-1ex]{0pt}{3.5ex}  HD~159222 & 5.1 & 141  & 1.0 & 31 & 24  \\
\hline 
\end{tabular}
\end{center}
\end{table}

\begin{figure} [ht]
   \begin{center}
   \begin{tabular}{c} 
   \includegraphics[height=10cm, trim=0 26cm 0 0]{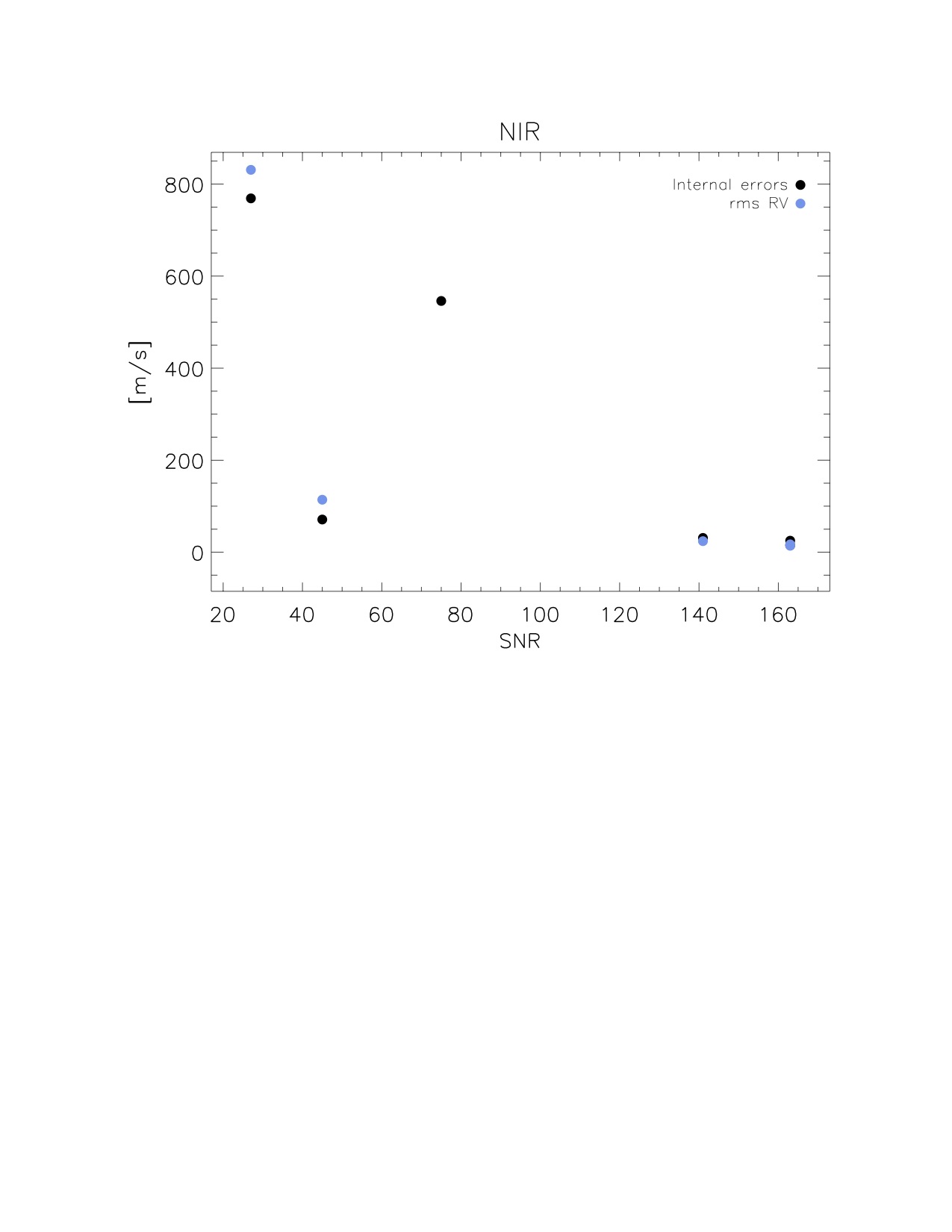}
	\end{tabular}
	\end{center}
   \caption[example] 
   { \label{fig:snr} 
Internal errors (red points) and RV r.m.s. scatter (cyan points) of our sample as function of the signal to noise ratio.}
   \end{figure}

\begin{figure} [ht]
   \begin{center}
   \begin{tabular}{c} 
   \includegraphics[height=10cm, trim=0 26cm 0 0]{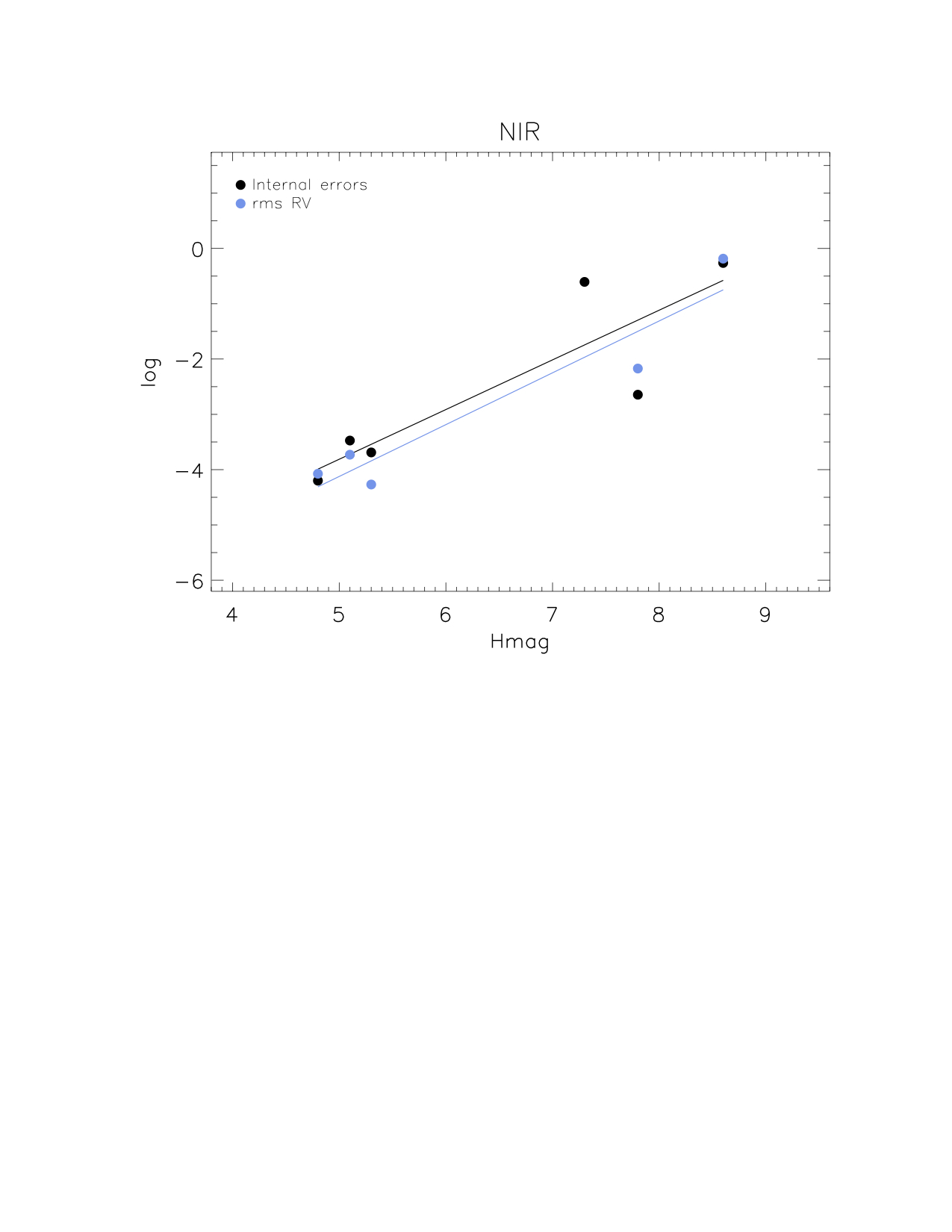}
	\end{tabular}
	\end{center}
   \caption[example] 
   { \label{fig:errnir} 
Logarithmic internal errors (red points) and RV r.m.s. scatter (cyan points) of our sample as function of the H magnitude.}
   \end{figure}

In Fig. \ref{fig:snr} the internal errors and the RV scatter are plotted as function of the SNR. The two highest internal errors correspond to the highest stellar $vsini$ values. Fig. \ref{fig:errnir} shows the same result in terms of H magnitude. The stars with an H magnitude of about 5 have a precision of 8 $ms^{-1}$ short-term and 14 $ms^{-1}$ long-term, while the stars with H magnitude of about 8 reach a precision of $~$100 $ms^{-1}$. 

The correlation, that we found between RV precisions and H magnitudes, can be represented with the following formula: $err_{RV}$ ($ms^{-1}$)=$10^{0.4 Hmag - 0.83}$.

\section{Future Works}
\label{sec:fut}
At the end of the 2017 Spring GIARPS has been delivered to the TNG in order to finish the instrument integration mainly with the Telescope Software environment. The main open points are the followings:
\begin{itemize}
  \item The GIANO--B command dispatcher: the GIANO--B NSTS version is under construction. At the moment it is possible to perform a couple of AB -- AB Nodding and use the spectrograph in stare observing mode. Most of the calibration templates are already built, other will follow also due to the future implementation of the Fabry perot system and the possible presence of absorbing cell.
  \item Absorbing Cells: The RVs from GIANO--B spectra are measured by using the CCF (cross-correlation function) method (see Section\ \ref{sec:rv}), with the telluric spectrum as wavelength reference.  The use of this technique allows a precision of about 8 m/s for bright stars (H$\leq 5$\ mag) and about 100 m/s for fainter stars (tipically H$\sim 9$ mag). The introduction of the absorption cell \cite{seemanetal2018} will allow more precise RV measurements with internal errors of about 3 m/s, due to the fact that absorption cell is more reliable in comparison with the instability of the telluric spectrum.
  \item RVs measurement: the GOFIO pipeline at the moment is limited to the on -- line production of completed reduced and wavelength calibrated GIANO--B spectra. In the next future, GOFIO will be able to automatically obtain the value of the Radial velocity from the single spectrum. Due to the used techniques to obtain RVs (both telluric or absorbing cell) GOFIO will process off -- line the spectra to extract the RV value  (at the end of the obesrving night).
  \item Fabry--Perot: at this moment the wavelength calibration is made by means of an U--Ne Lamp. A step further in the wavelength calibration precision (and so in the RV measurement) could be achieved with the larger number of lines given by a Fabry Perot System coupled with the calibration box of GIANO--B. The GIANO--B Fabry Perot system is described in Ref.\ \citenum{tozzi2016}.
\end{itemize}

\section{conclusions}
Since Fall 2017, GIARPS works routinely at the TNG that has a high-resolution spectroscopy station. Thank to its wide wavelength range (up to 2.5 $\mu$m) it is unique in the northern hemisphere and up to the commissioning of NIRPS (the NIR counterpart of HARPS) at the 3.6m ESO Telescope, the unique in this world. The flexibility of the three observing modes: HARPS-N alone, GIANO--B alone and GIARPS itself will allow users to select the best wavelength range useful for their preferred science case. From small bodies of the Solar System to the search for extrasolar planets will be the major science cases. 
Moreover it allows to reach a RV precision of 8m/s in the short-term, and 14m/s in the long-term in the NIR range. Furthermore, we presented the great contribution of the simultaneous observations to retract or confirm exoplantes orbiting active stars.  

\acknowledgments 
Authors would like to thank the support by INAF through Progetti Premiali funding scheme of the Italian Ministry of Education, University, and Research.

\bibliography{report} 
\bibliographystyle{spiebib} 

\end{document}